\documentclass[a4paper,fleqn]{cas-sc}

\usepackage[authoryear]{natbib}

\usepackage{multirow}
\usepackage{threeparttable}

\def\tsc#1{\csdef{#1}{\textsc{\lowercase{#1}}\xspace}}
\tsc{WGM}
\tsc{QE}

\begin{document}
\let\WriteBookmarks\relax
\def\floatpagepagefraction{1}
\def\textpagefraction{.001}

\shorttitle{Speaker and Language Diarization}    

\shortauthors{Anfeng Xu et al}  

\title [mode = title]{Speaker Role and Language Diarization for Analyzing Multilingual Interviews for Language Proficiency of Older Adults}  

\tnotemark[1] 

\tnotetext[1]{This document is the result of the research project funded by National Institutes of Health (NIH) under Grant Nos. R01 AG051125, U01 AG064948, RF1 AG055273, and R01AG080473.} 

%

\author[1]{Anfeng Xu}[orcid=0009-0001-5780-9797]
\cormark[1]
\ead{anfengxu@usc.edu}
\ead[url]{https://anfengxu136.github.io/}
\credit{Conceptualization, Methodology, Software, Formal analysis, Investigation, Data curation, Validation, Visualization, Writing -- original draft.}
\author[1]{Tiantian Feng}
\credit{Methodology, Data curation, Software, Writing -- review \& editing, }
\author[1]{Kevin Huang}
\credit{Investigation, Data curation, Writing -- review \& editing.}
\author[2]{Pranali Khobragade}
\credit{Data curation, Resources, Writing -- review \& editing.}
\author[1]{Sudarsana Kadiri}
\credit{Data curation, Resources, Writing -- review \& editing.}
\author[2]{Anushikha Dhankhar}
\credit{Data curation, Project administration, Writing -- review \& editing.}
\author[3]{Madeleine Snider}
\credit{Data curation, Project administration, Writing -- review \& editing.}
\author[2]{Sarah Gao}
\credit{Project administration, Resources, Writing -- review \& editing.}
\author[3]{Miguel Arce Rentería}
\credit{Supervision, Project administration, Resources, Writing -- review \& editing.}
\author[2]{Jinkook Lee}
\credit{Supervision, Project administration, Resources, Writing -- review \& editing.}
\author[1]{Shrikanth Narayanan}
\credit{Supervision, Project administration, Resources, Writing -- review \& editing.}


\affiliation[1]{organization={Viterbi School of Engineering, University of Southern California},
            addressline={3740 McClintock Ave}, 
            city={Los Angeles},
            postcode={90089}, 
            state={CA},
            country={USA}}

\affiliation[2]{organization={Center for Economic and Social Research, University of Southern California},
            addressline={635 Downey Way}, 
            city={Los Angeles},
            postcode={90007}, 
            state={CA},
            country={USA}}
            
\affiliation[3]{organization={Columbia University Irving Medical Center},
            addressline={622 W 168th St}, 
            city={New York},
            postcode={10032}, 
            state={NY},
            country={USA}}

\cortext[1]{Corresponding author}



\begin{abstract}
Automatic language proficiency assessment in the context of multilingual interview-based settings remains underexplored. In this work, we develop Whisper-based speaker-role and language diarization systems to automatically extract respondent speech and characterize language usage in multilingual interviews with older adults. We further investigate whether diarization-derived conversational and language-use behaviors can support downstream language proficiency assessment. Results show that language-adapted Whisper models substantially improve language diarization performance for lower-resource and linguistically related Indian languages. Statistical analyses reveal that respondent speech ratio and intended language usage are strong predictors of proficiency ratings. Furthermore, simple diarization-derived behavioral features achieve performance comparable to Whisper-based speech embeddings for proficiency prediction, while combining both yields the best results. Importantly, both the speech and language use statistical analyses and language proficiency prediction performance remain largely preserved when using fully automatic diarization outputs, demonstrating the potential of respondent-centric conversational analysis for scalable language proficiency assessment.
\end{abstract}


\begin{highlights}
\item We curate speech data from interviews with older adults in India.
\item We manually annotate the speech interviews dataset.
\item We develop Whisper-based speaker-role and language diarization systems.
\item Speech participation and intended-language use relate to language proficiency.
\item Automatic diarization yields proficiency predictions comparable to manual regions.
\end{highlights}


\maketitle
\section{Introduction}
Spoken language proficiency is a key indicator of communication competence and cognitive functioning, particularly among older adults \citep{roark2011spoken}. As populations continue to age worldwide, there is a growing interest in developing automatic methods for analyzing speech and language as scalable tools for clinical assessment, cognitive monitoring, and large-scale epidemiological studies. In multilingual settings, spoken language proficiency assessments are commonly used to evaluate an individual's ability to communicate effectively in one or more languages through measures of fluency, grammatical competence, and language use \citep{garcia2026proficiency}. However, most existing automatic cognitive and language proficiency assessment systems focus primarily on lexical, acoustic, or pronunciation-related features extracted from single-speaker recordings, while relatively little attention has been given to multi-speaker and multilingual scenarios.

Recent advances in speech foundation models~\citep{yang2024large} have substantially improved performance across a wide range of speech processing tasks, including automatic speech recognition (ASR)~\citep{prabhavalkar2023end}, speaker diarization~\citep{park2022review}, and language identification~\citep{o2025spoken}. These models provide rich contextualized speech representations that capture both acoustic and linguistic information. Nevertheless, language proficiency assessment in spontaneous conversational speech remains challenging. In interviewer--respondent interactions, the recorded audio contains speech from both participants, as well as pauses, nonverbal vocalizations, and occasional overlapping speech. As a result, obtaining accurate respondent speech regions is a critical prerequisite for downstream language analysis. Furthermore, speech behaviors such as the respondent speech ratio, mean length of utterance (MLU), intended language use, and nonverbal vocalizations may provide valuable indicators of language proficiency that are not explicitly captured by speech embeddings alone. Furthermore, in multilingual interview settings, particularly given the multiple linguistic backgrounds commonly found among individuals in India, respondents often possess varying degrees of proficiency across multiple languages and may engage in code-mixing or code-switching during spontaneous speech. As a result, language-use behaviors such as intended-language use may serve as informative indicators of proficiency, motivating the need for accurate language diarization in addition to respondent speech extraction.

To enable such analyses, automatic systems must first identify the portions of the conversation corresponding to the respondent and subsequently characterize language usage within those speech regions. While speaker role diarization~\citep{kim2025hybrid, xu2026end} and language diarization~\citep{kalluri2024second, mishra2023end, mishra2024implicit, chua2023merlion} have been studied as standalone speech processing tasks, their utility for downstream language proficiency assessments remains largely unexplored. In particular, it is unclear whether automatically extracted respondent speech regions and language usage patterns can provide reliable behavioral measurements for language proficiency analysis and prediction.

To address these gaps, we present a framework for respondent speech extraction, language diarization, and language proficiency prediction, using multilingual interview recordings collected as part of the Longitudinal Aging Study in India — Diagnostic Assessment of Dementia (LASI-DAD)~\citep{lee2020design}. The dataset consists of interviewer--respondent interactions collected during a spoken language proficiency assessment involving spontaneous storytelling tasks of multiple languages in India. We manually annotate speaker-role and language labels by annotators who are fluent in the corresponding languages used in individual interviews, enabling the study of respondent participation patterns and multilingual language usage behaviors. Building upon speech foundation models, such as Whisper~\citep{radford2023robust}, we develop a respondent speech extraction system implemented through speaker-role diarization, followed by a language diarization system that characterizes language usage within the extracted speech regions.

Beyond diarization, we have rated the language proficiency scores for each recording by expert raters proficient in the corresponding languages. We investigate how automatically derived conversational and language-use features relate to language proficiency. Using bootstrap-based statistical analyses, we examine the associations between respondent participation, utterance-level conversational patterns, language usage behaviors, and proficiency scores. We further evaluate language proficiency prediction using diarization-derived conversational features, acoustic-prosodic features, and speech foundation model embeddings. Through these analyses, we assess whether automatically inferred respondent speech regions and language labels preserve meaningful information for downstream proficiency assessment in older adults.

Our results demonstrate that conversational participation and language-use behaviors provide strong and interpretable indicators of language proficiency in older adults in India. In particular, respondent speech participation and intended-language usage exhibit consistent positive associations with proficiency ratings, and these relationships remain robust even when using fully automatic diarization outputs. Furthermore, simple diarization-derived conversational features achieve performance comparable to Whisper-based speech embeddings for language proficiency prediction in Hindi, while combining the two yields the strongest regression and classification results. These findings highlight the importance of respondent-centric conversational analysis and demonstrate the value of integrating speech foundation models with automatic respondent speech extraction and language diarization for scalable language proficiency assessment in multilingual older-adult populations.

The primary contributions of this work are as follows:

\begin{itemize}
\item We curate a multilingual older-adult interview-based speech dataset from India, with manually annotated speaker-role and language labels as well as spoken language proficiency ratings.

\item We develop and evaluate Whisper-based speaker-role diarization and language diarization systems for automatic respondent speech extraction and Indic language characterization.

\item We conduct statistical analyses of conversational and language-use behaviors and show that respondent speech ratio and intended-language usage are significant predictors of language proficiency scores, providing interpretable behavioral markers compared to relying on speech representations from deep learning models.

\item We evaluate language proficiency prediction in Hindi using diarization-derived features, acoustic-prosodic features, and Whisper encoder representations. We show that simple diarization-derived features provide strong, interpretable, and complementary information comparable to Whisper-based embeddings.

\item We demonstrate that both speech and language use statistical analyses and language proficiency prediction results remain largely consistent when using automatically inferred speaker-role and language labels rather than manual annotations, indicating that the proposed diarization systems preserve the key conversational and language-use information required for reliable downstream assessment.

\end{itemize}

\section{Related Works}
\subsection{Speech Processing for Automatic Spoken Language Proficiency Assessment}
Early speech-based automatic language proficiency assessment approaches primarily relied on automatic speech recognition (ASR) systems and handcrafted acoustic, prosodic, and linguistic features derived from speech and transcription outputs. For example, works such as \citep{wang2018towards, 8683268} developed an automatic proficiency assessment system using ASR-generated transcripts and neural network models, demonstrating the feasibility of predicting spoken language proficiency directly from spontaneous speech responses. Similarly, \citep{knill2018impact} studied the impact of ASR errors on automated spoken language assessment in question-answering tasks, where proficiency evaluation relied on linguistic features extracted from ASR-generated transcriptions. More recently, \citep{kang2024ai} developed large-scale AI-based language tutoring systems that leverage end-to-end ASR for non-native speech recognition and automatic fluency evaluation.

In parallel, recent work has increasingly adopted self-supervised speech representations to improve the performance over handcrafted features and reduce reliance on ASR-derived measurements, which are not always reliable under low-resource languages and populations. Works such as \citep{10023019, banno2023assessment, liu2023asr} explored the use of a self-supervised learning (SSL) based speech foundation model, wav2vec~2.0~\citep{baevski2020wav2vec}, for spoken language proficiency assessment and demonstrated that pretrained speech encoders capture proficiency-related information directly from acoustic signals. More recently, \citep{ma2025assessment} has explored Speech Large Language Models for L2 oral proficiency assessment, leveraging large-scale pretrained speech representations to directly model proficiency from spoken responses.

Despite these advances, most prior studies focus on isolated speech recordings from language learners and assume direct access to the target speaker's speech. In contrast, our work addresses multilingual interviewer--respondent interactions involving older adults, where the target respondent speech must first be extracted from conversational recordings before downstream language analysis can be performed. We further investigate how automatically derived conversational participation and language-use behaviors obtained through speaker-role and language diarization contribute to language proficiency assessment.

\subsection{Whisper as a Speech Foundation Model}
Recent work has demonstrated that Whisper can serve as a powerful speech foundation model whose learned representations encode diverse acoustic, linguistic, speaker, and behavioral characteristics beyond automatic speech recognition. In the large-scale Vox-profile benchmark, \citep{feng2025vox} systematically evaluated speech foundation models across a broad range of speaker and speech trait prediction tasks, including demographic attributes, vocal characteristics, and speech behaviors. Their results showed that Whisper representations consistently capture rich speaker- and speech-related information and achieve competitive or superior performance compared to other speech foundation models, highlighting Whisper's effectiveness as a general-purpose speech representation model.

Whisper has also shown strong performance in speaker diarization and speaker role diarization tasks. In child--adult conversational speech, \citet{xu2024exploring} demonstrated that Whisper-based representations substantially outperform traditional speaker diarization approaches such as Pyannote~\citep{plaquet2023powerset} and VBx~\citep{palka2026vbx}, while also outperforming self-supervised learning models such as Wav2Vec 2.0~\citep{baevski2020wav2vec} and WavLM~\citep{chen2022wavlm} for distinguishing child and adult speakers. Similar findings were reported by \citep{xu2026exploring}, which found that Whisper shows competitive diarization performance on older adults compared to DiariZen~\citep{han2025leveraging}, a more recent diarization system that leverages WavLM. 

Beyond speaker characterization, Whisper representations have also demonstrated strong capabilities for language modeling and clinical speech analysis. The VoxLect benchmark \citep{feng2026voxlect} introduced a large-scale evaluation framework for dialect and regional language identification across diverse languages worldwide and showed that language-adapted Whisper models consistently outperform SSL-based speech foundation models such as WavLM, highlighting the language-discriminative properties of Whisper embeddings. Similar trends have been observed in clinical speech analysis. In VoxCog \citep{feng2026voxcog}, which investigated multilingual cognitive impairment classification across diverse dialectal and linguistic conditions in older adults, Whisper-based models again outperformed SSL-based architectures, including WavLM. These findings suggest that Whisper representations capture not only language and dialect information but also clinically relevant speech and linguistic characteristics associated with cognitive decline, making them particularly suitable for downstream speech analysis tasks involving linguistically and demographically diverse populations.

Collectively, these studies demonstrate that Whisper functions as a versatile speech foundation model capable of modeling speaker traits, language identity, and cognitive characteristics. Motivated by these findings, we adopt Whisper as the backbone architecture for all stages of our framework, including speaker role diarization for respondent speech extraction, language diarization for multilingual language usage analysis, and Whisper-based language proficiency prediction.

\section{Dataset Curation}

\subsection{Data Collection and Statistics}

\begin{table}[t]
\centering
\begin{threeparttable}
\caption{Dataset statistics. Duration statistics are reported at the recording level.}
\label{tab:dataset_stats}
\begin{tabular*}{\linewidth}{@{\extracolsep{\fill}}lc@{}}
\toprule
\textbf{Statistic} & \textbf{Value} \\
\midrule

Number of Recordings & 548 \\
Number of Interviewer Speakers & 63 \\
Number of Respondent Speakers & 360 \\

\quad\quad Respondent Speaker Age Group$\dagger$ (years) & Mean$\pm$Std: 71.2$\pm$6.7, Range: 60-105, Median: 70 \\
\quad\quad  Respondent Speaker Gender$\dagger$ & 143 Men, 124 Women, 93 Not Available \\

Recording Duration (seconds) & Mean$\pm$Std: 63.8$\pm$31.2, Range: 20.8-344.7, Median: 65.1 \\
Audio Duration & \\
\quad\quad  Full Recordings & 9.71 hours \\
\quad\quad  Respondent Speech & 4.46 hours \\
\quad\quad  Interviewer Speech& 0.94 hours \\

\bottomrule
\end{tabular*}

\begin{tablenotes}
\footnotesize
\item[$\dagger$] Statistics are calculated based on 267 respondents with demographic information available.
\end{tablenotes}
\end{threeparttable}

\end{table}

The dataset is derived from the Wave-2 Objective Language Proficiency assessment within the Diagnostic Assessment of Dementia component of the Longitudinal Aging Study in India (LASI-DAD) \citep{lee2020introduction}. The task is designed to evaluate multilingual spoken language proficiency among older adults in India through spontaneous storytelling recordings. Multilingual participants are asked to narrate a familiar story in up to three reported languages, including stories such as \textit{Ramayana}, \textit{Mahabharata}, \textit{Akbar \& Birbal}, and other culturally familiar narratives. Alternatively, participants may describe a memorable festival or movie. The recordings are collected under realistic conversational and environmental conditions and are subsequently rated by language proficiency raters for fluency and grammatical competence on a five-point scale. 

The dataset contains multilingual conversational recordings involving an interviewer and an older adult respondent. A single participant may contribute multiple recordings if they reported proficiency in multiple languages. The recordings exhibit substantial variability in spontaneous speech patterns, conversational participation, language mixing, recording quality, and speaking proficiency.

Table~\ref{tab:dataset_stats} summarizes the dataset statistics. In total, the dataset contains 548 recordings from 360 speakers, corresponding to approximately 9.71 hours of audio. The respondents' speech accounts for approximately 4.46 hours, while the interviewers' speech accounts for approximately 0.94 hours. The participants are older adults aged 60 to 105 years ($71.2 \pm 6.7$ years), with a relatively balanced gender distribution. The average recording duration is 63.8 seconds with substantial variability due to differences in conversational fluency and response length.

\subsection{Annotations}

\subsubsection{Language Proficiency Rating}

\begin{table}[h]
\centering
\caption{Proficiency rating scale descriptors.}
\label{tab:fluency_rating}
\begin{tabular}{cp{14cm}}
\toprule
\textbf{Score} & \textbf{Description} \\
\midrule
1 & \textbf{Non-fluent with grossly incomplete sentences.} Speech is slow, hesitant, and strained except for short phrases; difficulty in perceiving continuity of speech; frequent grammatical errors even in simple structures; meaning is obscured. \\
2 & \textbf{Partially fluent with short sentences.} Speech is frequently hesitant, with some sentences left uncompleted; frequent grammatical errors even in simple structures that at times obscure meaning. \\
3 & \textbf{Vague and repetitive fluent speech.} Speech is relatively smooth; some hesitation and unevenness caused by rephrasing and searching for words; may have many repetitions; frequent grammatical errors that do not obscure meaning; little variety in grammatical structures. \\
4 & \textbf{Mostly complete and relevant sentences.} Smooth and fluid speech; few hesitations; a slight search for words; some errors in grammatical structures. \\
5 & \textbf{Sentences of adequate length and complexity.} Smooth and fluent speech; few to no hesitations; no attempts to search for words; accuracy and variety of grammatical structures. \\
\bottomrule
\end{tabular}
\end{table}

The recordings are evaluated by human annotators from an external transcription agency who are proficient in the corresponding languages. Each rater accesses the recordings through a dedicated rating platform and rates only languages within their expertise. The primary proficiency score is based on a five-point scale assessing fluency and grammatical competence. The rating guidelines are listed in Table~\ref{tab:fluency_rating}, which is a modified and abbreviated version of \citep{garcia2026proficiency}. In addition to the proficiency rating, raters also annotate recording quality, the presence of code-mixing and code-switching, and whether the respondent spoke in the intended target language.

Raters may skip recordings when reliable evaluation is not possible due to factors such as poor audio quality or the respondent primarily speaking a language other than the prompted language. In this work, we use the fluency and grammatical competence rating as the primary language proficiency score for downstream statistical analysis and automatic prediction experiments.

Table~\ref{tab:rating_stats} summarizes proficiency-score statistics for the overall dataset and the five most frequently represented languages. To assess rating reliability, co-authors familiar with Hindi, Telugu, Marathi, and English independently have rated 104 recordings (at least 25 per language) using the same five-point scale. We compare these ratings with the original ratings using Pearson correlation coefficient (PCC), quadratic-weighted kappa ($\kappa_w$), and within-1 agreement.
Overall, the ratings show moderate correspondence (PCC$=0.454$, $\kappa_w=0.303$, and 71\% within-1 agreement). Inter-rater agreement is strongest for Hindi and English, more modest for Marathi, and negligible for Telugu. Given the low inter-rater agreement for Telugu, we exclude Telugu from proficiency-based statistical and prediction analyses to avoid drawing conclusions from unreliable labels.

\begin{table}[h]
\centering
\caption{Proficiency rating statistics and inter-rater reliability.}
\label{tab:rating_stats}
\begin{tabular}{lccccccc}
\toprule
\textbf{Statistic} & \textbf{Hindi} & \textbf{Telugu} & \textbf{Marathi} & \textbf{English} & \textbf{Gujarati}  & \textbf{Others} & \textbf{All} \\
\midrule
Count & 158 & 99 & 60 & 57 & 42 & 122 & 538 \\[6pt]
Proficiency Rating (mean$\pm$std) & 2.6$\pm$1.1 & 2.4$\pm$0.6 & 2.5$\pm$0.8 & 2.9$\pm$1.2 & 2.9$\pm$0.8 & 2.8$\pm$1.0 & 2.6$\pm$1.0 \\[6pt]
Inter-rater Reliability &  &  &  &  &  &  &  \\
 - Pearson Correlation Coefficient (PCC) & 0.559 & 0.079 & 0.400 & 0.513 & N/A & N/A & 0.454 \\
 - Quadratic-Weighted Kappa ($\kappa_w$) & 0.391 & 0.049 & 0.159 & 0.417 & N/A & N/A & 0.303 \\
 - Within-1 agreement & 70\% & 72\% & 64\% & 78\% & N/A & N/A & 71\% \\
\bottomrule
\end{tabular}
\end{table}

\subsubsection{Speaker Role and Language Diarization Annotation}
We manually annotate both speaker role diarization and language diarization labels using Praat software. Each recording is annotated with three tiers: \textit{interviewer}, \textit{respondent}, and \textit{background-speech}. Annotators create utterance-level speech boundaries for conversational speech segments and assign corresponding speaker-role and language labels. 
To improve annotation consistency, we follow several temporal annotation constraints. Speech segments shorter than 0.1\,s are not annotated, while pauses shorter than 0.2\,s between speech from the same speaker are not segmented into separate utterances. Annotators are instructed to place speech boundaries with less than approximately 0.1--0.2\,s errors. For the interviewer and respondent tiers, each speech segment is assigned one of three label categories as detailed in Table~\ref{tab:label_categories}.

\begin{table}[h]
\centering
\caption{Label categories for interviewer and respondent speech segments.}
\label{tab:label_categories}
\begin{tabular}{ll}
\toprule
\textbf{Label Category} & \textbf{Description} \\
\midrule
Language ID & Intelligible vocalizations containing words, annotated using ISO 639-3 language identifiers. \\[6pt]
Unintelligible Speech & Speech containing words that are not intelligible. \\[6pt]
Nonverbal Vocalizations & Vocalizations without lexical content, such as laughter, ``uh-huh'', or filled pauses. \\
\bottomrule
\end{tabular}
\end{table}

\section{Method}

\begin{figure}
	\centering    \includegraphics[width=.8\textwidth]{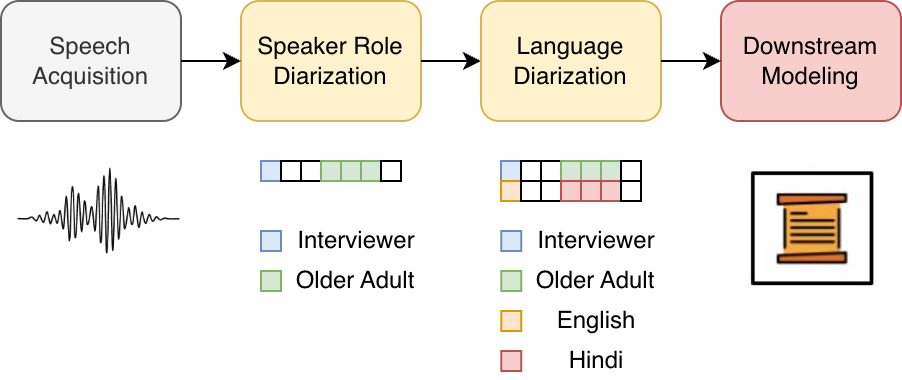}
	\caption{Illustration of speaker role diarization and language diarization pipelines.}
	\label{fig:pipeline}
\end{figure}

Fig.~\ref{fig:pipeline} presents the multi-stage computational pipelines, including speaker role diarization and language diarization modelings, followed by the downstream modeling. At the beginning, the dyadic conversational audio is processed by a speaker role diarization model to identify the speech segments associated with the respondent. Subsequently, language diarization is applied to the identified respondent speech segments to classify the spoken languages within each segment. We perform speaker role diarization before language diarization because we found it to be a more robust task, thereby minimizing error propagation in the downstream language diarization stage.

\subsection{Speaker Role Diarization Modeling for Respondent Speech Extraction}
We conduct respondent speech extraction via speaker role diarization modeling formulated as a frame-level classification problem. Specifically, the task is to classify each frame as interviewer, respondent, silence, or overlapping speech. Given an audio feature input $X = [x_1, ..., x_T]$, the goal is to predict the speaker-role label $y_t$ for $x_t$ at each timestamp $t$. Since this work is primarily developed based on the Whisper-family models, the input $X$ is defined as a 30-second audio feature segment with $T = 1500$ frames. This corresponds to the maximum frame length supported by the Whisper encoder at a frame shift of 20 ms.
Here,
$
y_t \in \{inter, res, sil, olp\},
$
where $inter$, $res$, $sil$, and $olp$ denote interviewer, respondent, silence, and overlapping speech, respectively.

The overall modeling pipeline is illustrated in Fig.~\ref{fig:pipeline}. We keep the Whisper encoder frozen and employ Low-Rank Adaptation (LoRA) on the feed-forward layers. The input waveform is first processed by the Whisper encoder to obtain hidden representations, followed by a learnable weighted average across encoder layers. The resulting features are passed through a stack of three 1D CNN layers, each with 256 channels, ReLU activation, and dropout with a probability of 0.1. A final 1D CNN layer with kernel size 1 produces frame-level predictions.

\subsection{Language Diarization Modeling}

Following the respondent speech extraction from speaker role diarization, we perform language diarization on the detected speech regions. Since language labels are only assigned to speech frames, silence regions are excluded from this stage.
Here, the target labels are 
$
y_t \in \{Hin, Eng, Tel, Mar, Guj, Other, Vo\},
$
where $Hin$, $Eng$, $Tel$, $Mar$, and $Guj$ denote Hindi, English, Telugu, Marathi, and Gujarati, respectively. The label $Other$ represents languages outside these target languages, while $Vo$ denotes nonverbal vocalizations.

We use the same Whisper-based architecture as described in Section~4.1. The only difference is the output space: the final prediction layer has four output channels for speaker role diarization and seven output channels for language diarization. During both training and inference, the entire audio segment is processed by the model. During training, the ground-truth annotation is used, and supervision is applied to both interviewer and respondent speech regions while silence, unintelligible speech, and overlapping speech regions are masked out for back-propagation. During inference, language predictions are generated for all frames, but only speech regions corresponding to the interviewer and respondent are retained, either from manual annotations or from the inferred speaker-role diarization outputs. For downstream statistical analysis and language proficiency prediction, only the respondent speech regions are used. For language diarization evaluation, overlapping speech and unintelligible speech regions are excluded from scoring.

\begin{figure}
	\centering
    \includegraphics[width=0.8\textwidth]{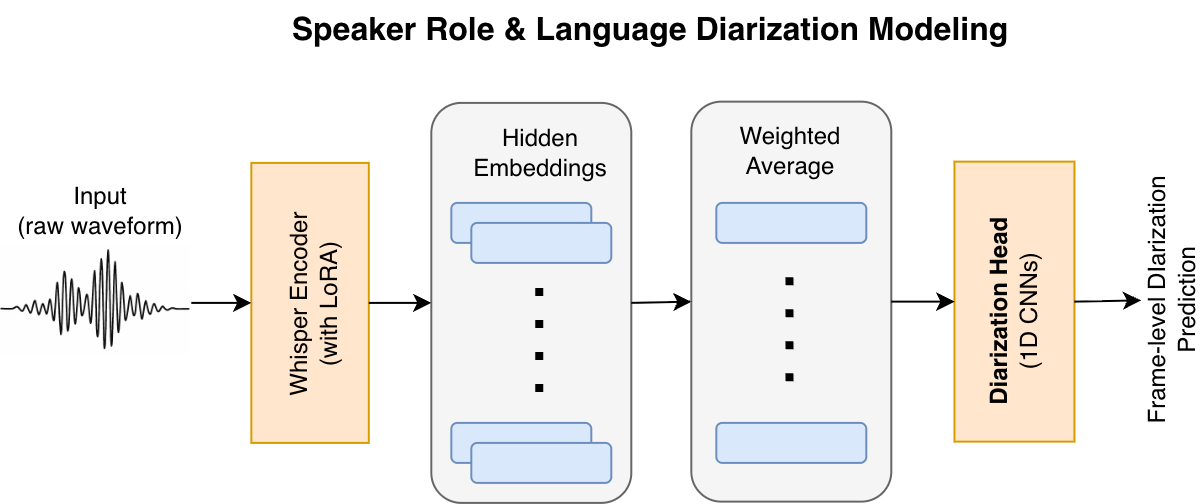}
	\caption{Illustration of speaker role and language diarization pipeline.}
	\label{fig:architecture}
\end{figure}

\subsection{Diarization Features and Statistical Analysis}

\label{sec:stat}

To analyze the relationship between diarization features and proficiency ratings, we perform bootstrap-based Ordinary Least Squares (OLS) regression analyses using both speaker role diarization and language diarization features extracted from the respondent speech segments. The dependent variable is the proficiency rating score, while the independent variables consist of normalized conversational features and language identity indicators. The extracted speaker role and language diarization features are listed in Table~\ref{tab:features}.

\begin{table}[h]
\centering
\caption{Extracted features for speaker diarization and language diarization analysis.}
\label{tab:features}
\begin{tabular}{llp{7cm}}
\toprule
\textbf{Analysis} & \textbf{Feature} & \textbf{Description} \\
\midrule
\multirow{5}{*}{Speaker Role Diarization}
    & Speech Ratio              & Total respondent speech duration divided by total audio duration. \\[4pt]
    & Utterance Length (mean)   & Mean duration of respondent utterances. \\[4pt]
    & Utterance Length (std)    & Standard deviation of respondent utterance durations. \\
\midrule
\multirow{4}{*}{Language Diarization}
    & Intended Language Ratio   & Proportion of intended-language speech within respondent utterances. \\[4pt]
    & Vocalization Ratio        & Proportion of nonverbal vocalizations within respondent utterances. \\
\bottomrule
\end{tabular}
\end{table}

For both speaker and language diarization analyses, we compare features extracted from ground-truth annotations and features derived from automatically inferred diarization outputs. This comparison evaluates the robustness of the statistical relationships under realistic automatic diarization conditions. The OLS model is formulated as:
\[
y = \sum_{i=1}^{M} \beta_i x_i + \sum_{k=1}^{K} \gamma_k \mathbb{1}\{\mathrm{lang}=k\} + \epsilon,
\]
where $y$ denotes the proficiency rating, $x_i$ represents the normalized diarization-derived features, $\beta_i$ and $\gamma_k$ are regression coefficients, $\epsilon$ is the residual error term, and $M$ and $K$ are the number of diarization features and the number of languages, respectively. Additionally, $\mathbb{1}\{\mathrm{lang}=k\}$ is the language indicator function that models language-specific offsets across language conditions. All continuous features are normalized prior to regression. For this experiment, we use the English, Hindi, and Marathi subsets, as these languages contain sufficient numbers of recordings, adequate variability in proficiency rating scores, and modest inter-rater agreement scores for reliable statistical analysis.

To assess coefficient stability and statistical significance, we conduct a bootstrap experiment with 10000 iterations. In each iteration, we randomly sample 80\% of the recordings from each language subset and fit the OLS model independently. We then compute the mean and standard deviation of each regression coefficient across bootstrap runs, along with the proportion of runs in which the corresponding $p$-value is below 0.05. We additionally report the mean adjusted $R^2$ across all bootstrap iterations.

\subsection{Language Proficiency Prediction}
We further investigate automatic prediction of language proficiency scores using three feature categories:
\begin{itemize}
    \item \textbf{Diarization-derived Features}: includes respondent speech ratio, utterance length (std), intended language usage, and vocalization ratio, as described in Section~\ref{sec:stat}. We exclude utterance length (mean), as it has shown no significance for language proficiency scores, as shown in Section~\ref{sec:result_prediction}.
    
    \item \textbf{OpenSMILE  Features}: handcrafted acoustic-prosodic features extracted using the OpenSMILE~\citep{eyben2010opensmile}. We use the eGeMAPSv02~\citep{eyben2015geneva} feature set consisting of 88 descriptors, capturing acoustic and prosodic characteristics such as pitch, energy, MFCC, jitter, shimmer, and harmonics-to-noise ratio.
    
    \item \textbf{Whisper Encoder Features}: speech representations obtained from a Whisper-small encoder from VoxLect, providing contextualized speech embeddings learned from large-scale speech pretraining.
\end{itemize}

Since Hindi contains the largest number of recordings, sufficient score variability, and relatively strong inter-rater agreement score, we conduct the prediction experiments on the Hindi subset. We evaluate both a regression task that directly predicts the continuous proficiency score on the 1--5 scale and a binary classification task separating low proficiency (scores 1--2) from higher proficiency (scores 3--5).

For the diarization-derived and OpenSMILE feature sets, we apply ridge regression for score prediction and logistic regression for binary classification. Both feature sets are computed using either oracle or automatically inferred respondent speaker regions. For inferred speaker regions, adjacent utterances separated by less than 0.2\,s are merged, while utterances shorter than 0.2\,s are discarded to reduce segmentation noise. The diarization-derived features are then computed from the resulting respondent speech regions. For the OpenSMILE-based approach, the processed respondent speech segments are concatenated prior to feature extraction. All handcrafted features are normalized prior to training.

For the Whisper-based approach, recordings are processed using sliding windows of up to 30\,s with a hop size of 15\,s during both training and inference. The waveform is passed through a Whisper-small encoder to obtain hidden representations, while the training objective is applied only to respondent speech regions and all other regions are masked out. We evaluate both oracle respondent regions and automatically inferred respondent regions obtained from the speaker diarization system. Audio segments shorter than 10\,s at the end of a recording are discarded. For binary classification, the model is trained using cross-entropy loss, while regression uses mean squared error (MSE) loss to directly estimate the 1--5 proficiency score.

We additionally evaluate a simple late-fusion ensemble strategy. For regression, the final predicted score is obtained by averaging the outputs from the individual models. For binary classification, we average the predicted class probabilities across models and assign the final label based on the resulting aggregated probability distribution.

\section{Experimental Settings}
\subsection{Metric for Speaker Role \& Language Diarization}
We evaluate both speaker role diarization and language diarization using the diarization error rate (DER). DER is computed with the following formula:
\begin{equation} \mathrm{DER} = \frac{\mathrm{MD} + \mathrm{FA} + \mathrm{Conf.}}{\mathrm{Total}}, \end{equation} where $\mathrm{MD}$ denotes missed speech, $\mathrm{FA}$ denotes false alarm speech,  $\mathrm{Conf.}$ denotes label confusion, and $\mathrm{Total.}$ denotes total reference speech duration. We use a collar tolerance of 0s around the utterance boundaries, and we include overlapping speech regions only for speaker role diarization evaluation, as those regions are masked out for language diarization. For speaker role diarization, the confusion term corresponds to speaker role confusion (SC), i.e., speech assigned to the wrong speaker role, such as interviewer versus respondent. Therefore, speaker role DER is: \begin{equation} \mathrm{DER}_{\mathrm{spk}} = \frac{\mathrm{MD} + \mathrm{FA} + \mathrm{SC}}{\mathrm{Total}}. \end{equation} For language diarization, the confusion term corresponds to language confusion (LC), i.e., speech assigned to the wrong language or vocalization category. Therefore, language DER is computed as: \begin{equation} \mathrm{DER}_{\mathrm{lang}} = \frac{\mathrm{MD} + \mathrm{FA} + \mathrm{LC}}{\mathrm{Total}}. \end{equation} For language diarization with oracle speech activity regions, $\mathrm{MD}$ and $\mathrm{FA}$ are zero by construction, and the DER is therefore equivalent to the language confusion rate.

\subsection{Data Splits and Cross-Validation} \label{sec:data_split} All experiments for diarization and language proficiency prediction are conducted using the same five-fold cross-validation split. The data are partitioned at the speaker level, ensuring that no respondent or interviewer appears in more than one fold. Furthermore, recordings from each language are distributed as evenly as possible across the five folds to maintain comparable language distributions throughout evaluation. For each cross-validation iteration, one fold is used for testing, while the remaining four folds are used for model development. Within the development portion, 25\% of the recordings are randomly selected as a validation set, with the remaining 75\% used for training. The same speaker-disjoint constraint is enforced between the training and validation sets.


\subsection{Experimental Details for  Speaker Role \& Language Diarization}

For speaker role diarization and language diarization, models are trained independently on each cross-validation
split. Evaluation is performed at the recording level, and the final reported metrics are obtained by averaging the test-
set results across the five folds. This procedure ensures that each recording is evaluated exactly once while providing
a robust estimate of overall performance.

For all diarization experiments, we train the models for 50 epochs using the Adam optimizer with a weight decay of $1\times10^{-4}$. We use a learning rate of $2\times10^{-3}$ for all diarization experiments, except for Whisper-VoxLect Large, for which we use a smaller learning rate of $5\times10^{-4}$, as higher learning rates result in unstable training.
During training, only the LoRA modules inserted into the feed-forward layers of the Transformer encoder and the diarization head are updated, while all other model parameters remain frozen. We use a LoRA rank of 64 for the Whisper-VoxLect variants, initializing both the LoRA parameters and the 1D CNN layers of the diarization head, except for the final output layer, from the corresponding VoxLect checkpoints. For the Whisper-Original variants, we use a smaller LoRA rank of 16, which we found to be more stable when training from randomly initialized LoRA parameters. We use a single NVIDIA RTX A6000 48GB GPU for all the experiments.

\subsection{Experimental Details for Language Proficiency Prediction}

For the language proficiency prediction experiments, we follow the same five-fold cross-validation protocol. Rather than averaging metrics independently across folds, we first aggregate the predictions from all test recordings across the five folds to form a single set of pooled out-of-sample predictions covering the entire dataset. The final evaluation metrics are then computed on this pooled prediction set. For experiments using inferred speaker and language diarization outputs, we use exactly the same train, validation, and test splits as the corresponding oracle experiments. The speaker role diarization and language diarization systems are first applied to all recordings within each split, and the resulting inferred boundaries and labels are used for subsequent feature extraction and model training. Consequently, both the training and evaluation stages operate entirely on automatically inferred diarization outputs, allowing us to assess the impact of diarization errors under a realistic deployment scenario while maintaining a fair comparison with the oracle condition.

For Whisper-based language proficiency prediction, we adopt the same architecture used for the diarization experiments, except that we apply temporal average pooling after the second CNN layer of the diarization head. The pooled representation is then passed to a linear output layer with a single output unit for both regression and binary classification. We use a learning rate of $2\times10^{-3}$, and all other training setups follow the same settings as for the diarization experiments.

\section{Results}
\subsection{Speaker Role \& Language Diarization Results}
\subsubsection{Speaker Role Diarization Results}

Table~\ref{tab:speaker_role_diarization} presents the speaker role diarization results using Whisper-based encoder representations. We compare Whisper-Original and Whisper-VoxLect initialization across Whisper-Small and Whisper-Large encoder configurations. Overall, all model configurations achieve very similar performance, with DER values ranging from 20.43\% to 21.62\%. The error patterns are also largely comparable across systems, with similar missed detection and false alarm rates. Speaker confusion remains relatively low for all configurations, ranging from 4.30\% to 5.77\%, indicating that the models are generally effective at distinguishing interviewer and respondent speech. These results suggest that both Whisper-Original and Whisper-VoxLect representations provide reliable speaker-role information for respondent speech extraction, with performance remaining stable across encoder scales and initialization conditions.

\begin{table}[h]
\centering
\caption{Speaker role diarization with different Whisper model families. The numbers are reported as percentages (\%)}
\label{tab:speaker_role_diarization}
\begin{tabular*}{\linewidth}{@{}LCCCCCCCC@{}}
\toprule
& \multicolumn{4}{L}{\textbf{Whisper-Small}} 
& \multicolumn{4}{L}{\textbf{Whisper-Large}} \\
\cmidrule(r){2-5}\cmidrule(r){6-9}
\textbf{Initialization} & \textbf{DER} & \textbf{MD} & \textbf{FA} & \textbf{SC}
& \textbf{DER} & \textbf{MD} & \textbf{FA} & \textbf{SC} \\
\midrule
Whisper-Original & 21.62 & 9.11 & 6.74 & 5.77 & 20.43 & 9.42 & 6.71 & 4.30 \\
Whisper-VoxLect & 20.76 & 8.50 & 7.26 & 5.00 & 21.01 & 8.35 & 7.68 & 4.98 \\
\bottomrule
\end{tabular*}
\end{table}

\subsubsection{Language Diarization Results}

\begin{table}[*h]
\centering
\caption{Language diarization results with different Whisper model families. The numbers are reported as percentages (\%).}
\label{tab:language_diarization}
\begin{tabular*}{\linewidth}{@{}LLCCCCCCCC@{}}
\toprule
& & \multicolumn{4}{L}{\textbf{Whisper-Small}} 
& \multicolumn{4}{L}{\textbf{Whisper-Large}} \\
\cmidrule(r){3-6}\cmidrule(r){7-10}
\textbf{Initialization} & \textbf{VAD} & \textbf{DER} & \textbf{MD} & \textbf{FA} & \textbf{LC}
& \textbf{DER} & \textbf{MD} & \textbf{FA} & \textbf{LC} \\
\midrule
\multirow{2}{*}{Whisper-Original} & Oracle & 47.71 & 0 & 0 & 47.71 & 35.03 & 0 & 0 & 35.03 \\
 & Predicted & 60.82 & 8.21 & 7.06 & 45.56 & 46.17 & 8.56 & 7.03 & 30.59 \\
\multirow{2}{*}{Whisper-VoxLect} & Oracle & 28.31 & 0 & 0 & 28.31 & 27.97 & 0 & 0 & 27.97 \\
 & Predicted & 40.38 & 7.62 & 7.56 & 25.20 & 40.47 & 7.49 & 8.05 & 24.93 \\
\bottomrule
\end{tabular*}
\end{table}

\begin{figure}
	\centering    \includegraphics[width=0.9\textwidth]{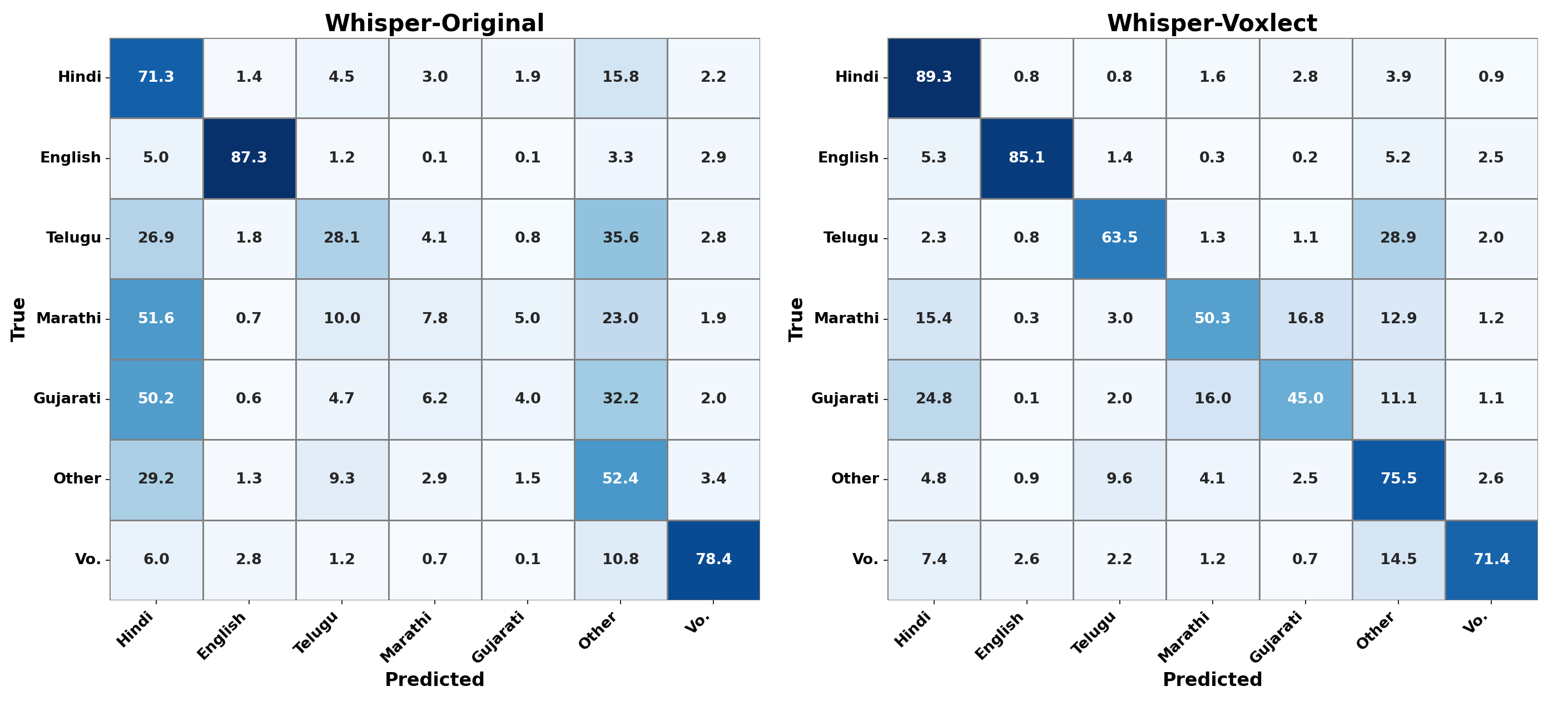}
	\caption{Confusion matrices for language diarization comparing Whisper-Original and Whisper-VoxLect using Whisper-Small. All numbers are reported in percentage (\%).}
	\label{fig:confusion_matrices}
\end{figure}

Table~\ref{tab:language_diarization} presents the language diarization results using Whisper-Small and Whisper-Large encoders under both Oracle and inferred VAD conditions. As expected, inferred VAD degrades performance relative to Oracle VAD for all systems due to additional missed detection (MD) and false alarm (FA) errors introduced by the speaker-role diarization stage. Overall, Whisper-VoxLect substantially improves language diarization performance across both model sizes and VAD conditions. Under Oracle VAD, the language confusion (LC) decreases from 47.71\% to 28.31\% for Whisper-Small and from 35.03\% to 27.97\% for Whisper-Large. Similar improvements are observed under inferred VAD, where LC decreases from 45.56\% to 25.20\% for Whisper-Small and from 30.59\% to 24.93\% for Whisper-Large. These consistent reductions indicate that language-identification-oriented adaptation produces substantially more language-discriminative representations for frame-level language attribution. Under the original Whisper initialization, Whisper-Large achieves lower LC than Whisper-Small, while after VoxLect adaptation, both model sizes achieve very similar performance. This suggests that the benefits of VoxLect adaptation are robust across encoder scales and are not limited to a particular model size. 

The confusion matrices in Figure~\ref{fig:confusion_matrices} provide additional insight into the performance gains from VoxLect adaptation. The largest improvements are observed for Telugu, Marathi, and Gujarati, which are among the most challenging language categories for the original Whisper model. For Telugu, recall increases from 28.1\% to 63.5\%, while confusion with Hindi decreases from 26.9\% to 2.3\%. Similarly, Marathi recall improves from 7.8\% to 50.3\%, accompanied by substantial reductions in confusion with Hindi (51.6\% to 15.4\%) and Other languages (23.0\% to 12.9\%). Gujarati exhibits a similar trend, with recall increasing from 4.0\% to 45.0\%, while confusion with Hindi decreases from 50.2\% to 24.8\% and confusion with Other languages decreases from 32.2\% to 11.1\%.

In contrast, languages that already achieve relatively strong performance with the original Whisper model, such as Hindi and English, exhibit smaller gains. Hindi recall improves from 71.3\% to 89.3\%, while English performance remains largely unchanged (87.3\% versus 85.1\%). The Other language category also benefits substantially, with recall increasing from 52.4\% to 75.5\%. Overall, these findings suggest that VoxLect adaptation is particularly effective at reducing confusion among closely related Indic languages, leading to substantially more discriminative language representations for lower-resource language categories.

\subsection{Statistical Analysis Results}

\subsubsection{Speaker Diarization Features}

\begin{table}[h]
\centering
\caption{OLS analysis using speaker diarization features. Oracle and Inferred denote annotation-derived and automatically inferred speaker-role regions, respectively. Coefficients are reported as mean $\pm$ standard deviation across runs.}
\label{tab:ols_speaker}
\small
\begin{tabular*}{\linewidth}{Lcc|cc}
\toprule

& \multicolumn{2}{c|}{\textbf{Oracle}}
& \multicolumn{2}{c}{\textbf{Inferred}} \\
\cmidrule(r){2-3}\cmidrule(l){4-5}

\textbf{Feature}
& \textbf{Coef.}
& \textbf{$P(p<0.05)$}
& \textbf{Coef.}
& \textbf{$P(p<0.05)$} \\

\midrule

Speech Ratio 
& 0.391$\pm$0.036 & 100.00\% & 0.368$\pm$0.037 & 100.00\% \\

Utterance Length (mean) 
& 0.009$\pm$0.054 & 0.01\% & -0.028$\pm$0.049 & 0.00\% \\

Utterance Length (std) 
& 0.034$\pm$0.054 & 0.03\% & 0.135$\pm$0.045 & 4.55\% \\

\midrule

\textbf{Adjusted $R^2$}
& \multicolumn{2}{c|}{0.153$\pm$0.021}
& \multicolumn{2}{c}{0.167$\pm$0.022} \\

\bottomrule
\end{tabular*}
\end{table}

Table~\ref{tab:ols_speaker} presents the bootstrap OLS analysis using speaker diarization features derived from both oracle annotations and automatically inferred speaker-role diarization outputs. We report the mean regression coefficients and the proportion of bootstrap runs in which each feature achieves statistical significance ($p<0.05$).

Across both oracle and inferred settings, speech ratio consistently emerges as the strongest predictor of language proficiency ratings. The feature exhibits stable positive coefficients ($0.391 \pm 0.036$ and $0.368 \pm 0.037$, respectively) and reaches statistical significance in 100\% of bootstrap runs under both conditions. This finding suggests that respondents who contribute a larger proportion of the interview conversation tend to receive higher proficiency ratings.

In contrast, mean utterance length shows coefficients close to zero and is almost never statistically significant, indicating little association with language proficiency in this dataset. Utterance length variability exhibits a modest positive relationship with proficiency. Under oracle speaker regions, the coefficient remains small ($0.034 \pm 0.054$) and is significant in only 0.03\% of bootstrap runs. When inferred speaker regions are used, the coefficient increases to $0.135 \pm 0.045$, although the feature still achieves statistical significance in only 4.55\% of bootstrap runs. These results indicate that utterance-level duration statistics contribute substantially less information than overall conversational participation.

The overall regression performance remains highly consistent between oracle and automatically inferred speaker regions. The adjusted $R^2$ is $0.153 \pm 0.021$ for oracle annotations and $0.167 \pm 0.022$ for inferred speaker regions, indicating that the automatically derived respondent speech segments preserve the primary conversational patterns associated with language proficiency. Overall, these findings suggest that respondent participation is the dominant conversational predictor of proficiency ratings, while the proposed speaker-role diarization system provides sufficiently reliable respondent speech extraction for downstream behavioral analyses.

\begin{table}[h]
\centering
\caption{OLS analysis using language diarization features. Oracle and Inferred denote annotation-derived and automatically inferred language features, respectively. Coefficients are reported as mean $\pm$ standard deviation across runs.}
\label{tab:ols_language}
\small
\begin{tabular*}{\linewidth}{Lcc|cc|cc}
\toprule
& \multicolumn{2}{c|}{\textbf{Oracle}} 
& \multicolumn{2}{c|}{\textbf{Inferred (Oracle VAD)}}
& \multicolumn{2}{c}{\textbf{Inferred (Predicted VAD)}} \\
\cmidrule(r){2-3}\cmidrule(l){4-5}\cmidrule(l){6-7}

\textbf{Feature} 
& \textbf{Coef.} 
& \textbf{$P(p<0.05)$}
& \textbf{Coef.} 
& \textbf{$P(p<0.05)$}
& \textbf{Coef.} 
& \textbf{$P(p<0.05)$} \\

\midrule

Intended Language Prob. 
 & 0.341$\pm$0.031 & 100.00\% & 0.235$\pm$0.037 & 96.81\% & 0.246$\pm$0.039 & 97.42\% \\

Nonverbal Vocalization Prob. 
& -0.144$\pm$0.035 & 35.54\% & -0.266$\pm$0.034 & 99.93\% & -0.212$\pm$0.042 & 94.24\% \\

\midrule

\textbf{Adjusted $R^2$}
& \multicolumn{2}{c|}{0.160$\pm0.020$}
& \multicolumn{2}{c|}{0.149$\pm0.022$}
& \multicolumn{2}{c}{0.124$\pm0.021$} \\

\bottomrule
\end{tabular*}
\end{table}

\subsubsection{Language Diarization Features}
Table~\ref{tab:ols_language} presents the bootstrap OLS analysis using language diarization features under three conditions: oracle annotations, inferred language diarization with oracle VAD, and fully inferred language diarization with predicted VAD obtained from speaker role diarization. The reported coefficients correspond to the mean and standard deviation across 10,000 bootstrap iterations.

Across all experimental settings, the intended language ratio exhibits a consistent positive association with language proficiency ratings. The feature achieves statistical significance in nearly all bootstrap runs, with significance rates of 100.0\%, 96.8\%, and 97.4\% for the oracle, inferred language diarization with oracle VAD, and fully inferred settings, respectively. The corresponding regression coefficients remain positive across conditions ($0.341 \pm 0.031$, $0.235 \pm 0.037$, and $0.246 \pm 0.039$), indicating that respondents who spend a larger proportion of their speech in the intended target language tend to receive higher proficiency ratings. Although the coefficients decrease when inferred language labels are used, the overall relationship remains highly stable, suggesting that intended language usage is a robust behavioral marker of language proficiency even under automatic diarization conditions.

Nonverbal vocalization ratio demonstrates a consistent negative relationship with proficiency ratings. Under oracle annotations, the feature exhibits a modest negative coefficient ($-0.144 \pm 0.035$) and reaches statistical significance in 35.5\% of bootstrap runs. Interestingly, the relationship becomes substantially stronger under automatic language diarization, with coefficients of $-0.266 \pm 0.034$ and $-0.212 \pm 0.042$ and significance rates of 99.9\% and 94.2\% for the oracle-VAD and predicted-VAD settings, respectively. These results suggest that a higher proportion of nonverbal vocalizations is associated with lower language proficiency. One possible explanation for the stronger effect under inferred language diarization is that less proficient speakers tend to produce more disfluent speech. These regions may receive higher nonverbal vocalization probabilities from the language diarization model, whereas human annotators may be more likely to label them as intended-language speech. As a result, the inferred vocalization ratio may capture broader speech disfluency patterns beyond true nonverbal vocalizations, leading to a stronger association with language proficiency.

The overall explanatory power of the regression model decreases as diarization errors are introduced, with adjusted $R^2$ values declining from $0.160 \pm 0.020$ for oracle annotations to $0.149 \pm 0.022$ and $0.124 \pm 0.021$ for the inferred language diarization conditions. Nevertheless, both intended language usage and nonverbal vocalization behaviors retain consistent directional relationships with proficiency ratings across all settings. These findings indicate that language-use patterns extracted from automatic speaker role and language diarization preserve the key behavioral signals required for downstream proficiency-related analyses.

\subsection{Language Proficiency Prediction}
\label{sec:result_prediction}

\begin{table}[h]
\centering
\caption{Linear regression results. Oracle and Inferred denote annotation-derived and automatically inferred speaker-role and language regions, respectively. PCC denotes Pearson Correlation Coefficient. }
\label{tab:regression}
\small
\begin{tabular*}{\linewidth}{LCCCC}
\toprule

& \multicolumn{2}{c}{\textbf{PCC}}
& \multicolumn{2}{c}{\textbf{MAE}} \\
\cmidrule(){2-3}\cmidrule(){4-5}

\textbf{Feature}
& \textbf{Oracle}
& \textbf{Inferred}
& \textbf{Oracle}
& \textbf{Inferred} \\

\midrule

Diarization-derived Features
& 0.441 & 0.440 & 0.794 & 0.806 \\

OpenSMILE Features
& 0.177 & 0.175 & 0.996 & 1.000 \\

Whisper Encoder Features
& 0.528 & 0.527 & \textbf{0.732} & \textbf{0.725} \\
\cmidrule{1-1}

Ensemble (Diarization + Whisper)
& \textbf{0.531} & \textbf{0.530} & 0.744 & 0.747 \\

\bottomrule
\end{tabular*}
\end{table}

\begin{table}[t]
\centering
\caption{Classification analysis. Oracle and Inferred denote annotation-derived and automatically inferred speaker-role and language regions, respectively. The results are shown in percentages.}
\label{tab:classification}
\small
\begin{tabular*}{\linewidth}{LCCCC}
\toprule

& \multicolumn{2}{c}{\textbf{Accuracy (\%)}}
& \multicolumn{2}{c}{\textbf{F1-Macro (\%)}} \\
\cmidrule(){2-3}\cmidrule(){4-5}

\textbf{Feature}
& \textbf{Oracle}
& \textbf{Inferred}
& \textbf{Oracle}
& \textbf{Inferred} \\

\midrule

Diarization-derived Features
& 61.39 & 62.03 & 61.39  & 62.00 \\

OpenSMILE Features
& 53.80 & 54.43 & 53.75 & 54.42 \\

Whisper Encoder Features
& 59.49 & 59.49 & 59.47 & 59.47 \\

\cmidrule{1-1}

Ensemble (Diarization + Whisper)
& \textbf{64.56} & \textbf{65.82} & \textbf{64.55} & \textbf{65.80} \\

\bottomrule
\end{tabular*}
\end{table}

Tables~\ref{tab:regression} and~\ref{tab:classification} present the language proficiency prediction results for the regression and binary classification tasks, respectively, using diarization-derived conversational features, OpenSMILE acoustic-prosodic features, Whisper encoder representations. We also report ensemble results combining predictions from diarization-derived and Whisper encoder features.

For the regression task in Table~\ref{tab:regression}, the Whisper encoder features achieve the strongest individual performance, obtaining PCC values of 0.528 and 0.527 under oracle and inferred speaker regions, respectively, while also achieving the lowest MAE of 0.732 and 0.725, respectively. The diarization-derived conversational features perform competitively, reaching PCC values of 0.441 and 0.440, despite relying on only a small set of high-level behavioral features. In contrast, the OpenSMILE feature set performs substantially worse, achieving PCC values below 0.18 and MAE values of around 1.00. The relatively strong performance of the diarization-derived features suggests that conversational participation patterns and language-use behaviors contain substantial information relevant to language proficiency assessment.

Combining the diarization-derived and Whisper-based systems through simple prediction averaging yields the strongest overall regression performance. The ensemble achieves PCC values of 53.12\% and 53.01\% under oracle and inferred speaker regions, respectively, while maintaining low MAE values of 0.74 and 0.75. Although the improvements over the Whisper-only system are modest, the consistent gains indicate that the diarization-derived behavioral features provide complementary information beyond the acoustic and linguistic representations captured by Whisper encoder embeddings. In particular, the behavioral features encode higher-level conversational dynamics and language-use statistics that are not explicitly represented in the speech embeddings.

For the binary classification task in Table~\ref{tab:classification}, the overall trends differ somewhat from the regression results. The diarization-derived features emerge as the strongest individual feature set, achieving accuracies of 61.39\% and 62.03\% and F1-Macro scores of 61.39\% and 62.00\% under oracle and inferred speaker regions, respectively. These results outperform both the Whisper encoder features and the OpenSMILE features, further highlighting the importance of conversational participation and language-use behaviors for distinguishing lower-proficiency speakers from higher-proficiency speakers. The ensemble model again achieves the best overall performance, reaching accuracies of 64.56\% and 65.82\% and F1-Macro scores of 64.55\% and 65.80\% under oracle and inferred speaker regions, respectively.

Across both regression and classification tasks, the performance differences between oracle and inferred speaker regions remain relatively small. For example, the regression ensemble achieves nearly identical PCC and MAE values under both conditions, while the classification ensemble exhibits slightly higher performance under inferred speaker regions. These findings suggest that the proposed speaker-role and language-diarization systems preserve the key conversational and language-use information required for downstream language proficiency prediction, making the overall pipeline practical for real-world deployment without requiring manual annotations.

Compared with the regression task, the performance gains from ensembling are more pronounced for binary classification. While the regression ensemble provides only marginal improvements over the Whisper-based system, the classification ensemble improves F1-Macro by approximately 3--6 percentage points over the individual feature sets. This result suggests that the diarization-derived behavioral features and Whisper encoder representations provide complementary decision boundaries for separating lower- and higher-proficiency speakers. In contrast, the regression task appears to be dominated by the Whisper representations, leaving less room for additional gains from behavioral feature fusion.

\section{Conclusion}

In this work, we have investigated automatic language proficiency assessment in multilingual interviewer–respondent interactions among older adults. We have developed Whisper-based speaker role diarization and language diarization systems to automatically extract respondent speech and characterize language usage patterns from conversational recordings. Using a multilingual subset of the LASI-DAD language proficiency assessment corpus, we have demonstrated that language-adapted Whisper models substantially improve language diarization performance, particularly for lower-resource and linguistically related languages such as Marathi and Gujarati.

Beyond diarization performance, we have shown that automatically derived conversational and language-use behaviors provide meaningful and interpretable indicators of language proficiency. In particular, the respondent speech ratio and intended-language usage showed strong, consistent positive associations with proficiency scores. Furthermore, simple diarization-derived behavioral features achieved performance comparable to Whisper-based speech embeddings for proficiency prediction, while combining behavioral and speech representation features yielded the strongest overall performance. Importantly, both the statistical relationships and prediction performance remained largely preserved when using fully automatic speaker-role and language diarization outputs, demonstrating the feasibility of scalable respondent-centric language proficiency assessment without manual annotations.

Future work can integrate ASR systems and speech LLMs into multilingual interview-based assessments to directly model lexical, grammatical, and semantic information in respondents’ speech. Combined with diarization-derived features and Whisper encoder representations, these approaches could provide a more holistic assessment of proficiency by jointly leveraging acoustic, linguistic, and interactional cues while accommodating multilingual speech, code-switching, and cross-lingual variation. Future studies could also compare speech behaviors when respondents use their first versus non-first languages and examine how language background affects the relationship between these behaviors and proficiency.





\clearpage 




\section{}\label{}

\printcredits

\bibliographystyle{cas-model2-names}

\bibliography{cas-refs}



\end{document}